\documentclass[pra,twocolumn,showpacs,,superscriptaddress,nofootinbib,floatfix]{revtex4}

\usepackage{graphicx}  
\usepackage{amsmath}
\usepackage{amsfonts}
\usepackage{bm}  
\usepackage{color}

\newcommand{\bea}{\begin{eqnarray}}
\newcommand{\eea}{\end{eqnarray}}

\newcommand{\be}{\begin{equation}}
\newcommand{\ee}{\end{equation}}
\newcommand{\bqa}{\begin{eqnarray}}
\newcommand{\eqa}{\end{eqnarray}}

\newcommand{\Rvec}{{\bm R}}
\newcommand{\rvec}{{\bm r}}
\newcommand{\kvec}{{\bm k}}

\newcommand{\qvec}{{\bm q}}
\newcommand{\pvec}{{\bm p}}
\def\mqo2{{\!\!\!}}

\newcommand{\cC}{\mathcal{C}}
\newcommand{\cD}{\mathcal{D}}
\newcommand{\cO}{\mathcal{O}}

\newcommand{\cI}{\mathcal{I}}
\newcommand{\cT}{\mathcal{T}}

\newcommand{\cF}{\mathcal{F}}
\newcommand{\cL}{\mathcal{L}}
\newcommand{\cA}{\mathcal{A}}

\newcommand{\cE}{\mathcal{E}}

\newcommand{\eq}[1]{Eq.~\eqref{eq:#1}}
\newcommand{\eqs}[2]{Eqs.~\eqref{eq:#1} and \eqref{eq:#2}}
\newcommand{\eqss}[3]{Eqs.~\eqref{eq:#1}, \eqref{eq:#2}, and \eqref{eq:#3}}
\renewcommand{\sec}[1]{Sec.~\ref{sec:#1}}

\newcommand{\fig}[1]{Fig.~\ref{fig:#1}}

\newcommand{\tab}[1]{Table~\ref{tab:#1}}
\newcommand{\nn}{\nonumber}

\newcommand{\Alo}{\mathcal{A}_\text{LO}} 
\newcommand{\Anlo}{\mathcal{A}_\text{NLO}}

\newcommand{\arrowpartial}[1]{\overleftrightarrow{\partial}\!\!_{#1}}
\newcommand{\larrowpartial}[1]{\overleftarrow{\partial}\!\!_{#1}}
\newcommand{\rarrowpartial}[1]{\overrightarrow{\partial}\!\!_{#1}}
\newcommand{\arrowpartials}[2]{\overleftrightarrow{\partial}\!\!_{#1}\overleftrightarrow{\partial}\!\!_{#2}}
\newcommand{\arrowpartialss}[3]{\overleftrightarrow{\partial}\!\!_{#1}\overleftrightarrow{\partial}\!\!_{#2}\overleftrightarrow{\partial}\!\!_{#3}}
\newcommand{\arrowpartialsq}{\overleftrightarrow{\nabla}^2}

\newcommand{\Alam}{\cA_\lambda}
\newcommand{\Arho}{\cA_\rho}
\newcommand{\Arhop}{\cA_{\rho'}}
\newcommand{\e}{\epsilon}

\newcommand{\Oc}{\psi_1^\dagger\psi_2^\dagger\psi_2\psi_1} 
\newcommand{\Od}{\psi_1^\dagger\psi_2^\dagger\psi_2 \overleftrightarrow{\nabla}^2 \psi_1}

\newcommand{\psis}{\psi_\sigma}
\newcommand{\psia}{\psi_1}
\newcommand{\psib}{\psi_2}
\newcommand{\psisdagg}{\psi_\sigma^\dagger}
\newcommand{\psiadagg}{\psi_1^\dagger}
\newcommand{\psibdagg}{\psi_2^\dagger}

\newcommand{\IaD}[1]{\cI^{(1,\Delta)}_{#1}}
\newcommand{\IbD}[1]{\cI^{(2,\Delta)}_{#1}}
\newcommand{\Ia}[2]{\cI^{(1,#1)}_{#2}}
\newcommand{\Ib}[2]{\cI^{(2,#1)}_{#2}}

\begin{document}
\preprint{LA-UR-15-29631}
\title{
The Operator Product Expansion Beyond Leading Order \\
for Two-Component Fermions}

\author{Samuel B. Emmons}\email{semmons@vols.utk.edu}
\affiliation{Department of Physics and Astronomy, University of
  Tennessee, Knoxville, TN 37996, USA}

\author{Daekyoung Kang}\email{kang1@lanl.gov}
\affiliation{Theoretical Division, MS B283,
Los Alamos National Laboratory, Los Alamos, NM 87545, USA }
\author{Lucas Platter}\email{lplatter@utk.edu} 
\affiliation{Department of Physics and Astronomy, University of
  Tennessee, Knoxville, TN 37996, USA}

\affiliation{Physics Division, Oak Ridge National Laboratory, Oak
  Ridge, TN 37831, USA}


\begin{abstract}
  We consider a homogeneous, balanced gas of strongly interacting
  fermions in two spin states interacting through a large scattering
  length.  Finite range corrections are needed for a quantitative
  description of data which experiments and numerical simulations have
  provided. We use a perturbative field theoretical framework and a
  tool called the Operator Product Expansion (OPE), which together
  allow for the expression of finite range corrections to the
  universal relations and momentum distribution.  Using the OPE, we
  derive the $1/k^6$ part of the momentum tail, which is related to
  the sum of the derivative of the energy with respect to the finite
  range and the averaged kinetic energy of opposite spin
  pairs. By comparing the $1/k^4$ term and the $1/k^6$ correction in
  the momentum distribution to provided Quantum Monte Carlo (QMC)
  data, we show that including the $1/k^6$ part offers marked
  improvements. Our field theoretical approach allows for a clear
  understanding of the role of the scattering length and finite
  effective range in the universal relations and the momentum
  distribution.
\end{abstract}


\smallskip
\maketitle
\newpage

\section{Introduction}
\label{sec:introduction}
Strongly interacting systems of ultracold two-component fermions have
been studied using various techniques for many years.
In nuclear physics this system is of interest
due to its simplicity and similarity to a gas of ultracold
neutrons. In atomic physics, this system is of interest because of its
transition from a Bose-Einstein condensate (BEC) at small positive
scattering lengths to a system that displays BCS superfluidity at small
negative scattering lengths. Specifically,
the case called the \textit{unitary limit}, in which the two-body
scattering length $a$ is taken to infinity, received a lot of
attention from experimentalists and theorists alike
\cite{Inguscio:2007cma}. In this limit the zero-range model can be
used to describe systems with large scattering length. In this model,
the range of the atom-atom interaction is taken to zero while the
binding energy is kept constant by adjustment of the coupling
strength. 

Universal relations, which are independent of the structure of the
particles and the state of the system, were derived in the zero range
limit for these systems by Shina Tan in 2005
\cite{2008AnPhy.323.2952T,2008AnPhy.323.2971T,2008AnPhy.323.2987T}. These
relations contain the so-called \textit{contact} that can be defined
as the asymptote of the $1/k^4$ large momentum tail of the momentum
distribution. The contact is a state dependent quantity and will
therefore depend on quantities such as the scattering length,
temperature, or density of the system.

Tan's contact and the related universal relations can be derived by
applying the operator product expansion
(OPE)~\cite{Wilson:1969zs,Kadanoff:1969zz,Polyakov:1970}, which is the
quantum field theoretical short-distance expansion of a nonlocal
operator. In Ref.~\cite{Braaten:2008uh}, it was shown that this tool
can be used to derive Tan's universal relations and that the contact
is related to the leading two-body interaction term in the OPE.  The
OPE has since then been used not only to derive additional universal
relations for the two-component Fermi gas, but it has also been applied to
novel systems such as the unitary Bose gas
\cite{Braaten:2008bi,Son:2010kq,Goldberger:2010fr,Goldberger:2011hh,
  Hofmann:2011qs,Braaten:2011sz,Hofmann:2012np,Nishida:2011xb,
  2013JPhB...46u5203B,Gubler:2015iva,Braaten:2010dv,Smith:2013eoa}
(for a discussion of other approaches that have been used to derive
universal relations see Ref.~\cite{Braaten:2010if} and the references
therein).

In this manuscript, we derive improved universal relations which
include the finite effective range of the two-body interaction. Just
as the contact was identified as playing an important role in the
zero-range limit, we identify two quantities that appear in
universal relations valid beyond the zero-range limit. One of them, which we
call the \emph{derivative contact}, is a measure of the sensitivity of
the energy of the system to the effective range.  The other measures
the averaged kinetic energy of opposite spin pairs at zero relative distance.
Some of these relations were already derived using a quantum mechanical framework by
Castin and Werner in Ref.~\cite{2012PhRvA..86a3626W}. Here we will use
the OPE framework and an effective field theory (EFT) to derive
additional finite-range universal relations. In the EFT approach an existing
separation of scales is turned into an expansion parameter for a
systematic low-energy perturbative expansion. In our case this expansion parameter
is the ratio $\ell/a$, where $\ell$ denotes the range of the atom-atom
interaction. This approach is very powerful since it makes no
assumptions regarding the microscopic interaction responsible for the
large scattering length and is therefore completely
model-independent.

First, we review the theory that is used to describe particles
interacting through a short-range interaction in
Sec.~\ref{sec:effect-field-theory}. We present the renormalization of
the theory up to both Leading Order (LO) and Next-to-Leading Order
(NLO) in the EFT expansion. Renormalization eliminates divergences
which arise in the field theoretical calculations and leaves us with
physical, finite results. After presenting the EFT model we use, we
move directly to the results and leave the detailed calculations of
those results for later.  Thus, Sec. \ref{sec:rels} shows universal
relations with effective range corrections, including a subleading,
in the OPE expansion, tail of the momentum distribution and
corrections to the energy, adiabatic, and pressure relations and to
the virial theorem. Section \ref{sec:result} gives a numerical
comparison between the contact and the derivative contact and shows
the OPE result in comparison to Quantum Monte Carlo (QMC) data.
Lastly, Sec. \ref{sec:OPE} contains the many details of the OPE
calculations which lead to the results already presented in
Sec. \ref{sec:rels}. The renormalized two-body operators in the OPE
form the contact and derivative contact operators which are in the
universal relations.

\section{Effective field theory}
\label{sec:effect-field-theory}
For an interaction with finite range, $\ell$, the two-body t-matrix
can be written as
\begin{equation}
  \label{eq:tmatrix-general}
  t(k)=\frac{4\pi}{m}\frac{1}{k\cot\delta_{0}-i k}~,
\end{equation}
where $m$ denotes the particle mass, $k$ is the relative momentum
between the two particles, and $\delta_{0}$ is the scattering
phaseshift. At sufficiently low energies, we can expand
$k\cot \delta_{0}$ using the effective range expansion
\begin{equation}
\label{eq:phaseshift}
  k\cot \delta_{0} =-\frac{1}{a}+\frac{r_s}{2}k^2+\ldots~,
\end{equation}
where $a$ is the $S$-wave scattering length, and $r_{s}$ is the
$S$-wave effective range. For $k \ell\ll 1$, the short-range details
of the interaction are not resolved and such systems can therefore be
described with an EFT that employs only contact interactions. 

The EFT Lagrangian for particles interacting through contact
interactions can be written as
\begin{align}
  \cL=\cL_0+\cL_1+\cdots
  \,,\end{align}
where $\cL_{0,1}$ are leading order (LO) and next-to-leading order
(NLO) in $r_s$ Lagrangians and the dots denote operators with more
derivatives and/or fields that contribute to higher orders in the EFT
expansion. The LO and NLO Lagrangians are given by
\begin{align}
\label{eq:L0}
\cL_0 &=
\sum_{\sigma=1,2}\psi_\sigma^\dagger\left(i\partial_t+\frac{\nabla^2}{2m}\right)\psi_\sigma
-\frac{\lambda_{0}}{m}\psi_1^\dagger\psi_2^\dagger\psi_2\psi_1
\\ 
\label{eq:L1}
\cL_1 &=
\frac{\rho_0}{4}\left(\psi_1^\dagger\psi_2^\dagger\psi_2\arrowpartialsq \psi_1+\rm{h.c.}\right) +\delta \cL_1
\,,\\
\label{eq:dL1}
 \delta \cL_1 &= -\frac{\delta\lambda_{0}}{m}\psi_1^\dagger\psi_2^\dagger\psi_2\psi_1
\,. \end{align}
Note that we have set $\hbar=1$.
When considered by itself, Lagrangian $\cL_0$ is also known as the
zero-range model, and we will show below how the bare coupling
$\lambda_0$ is related to the scattering length $a$ through
renormalization. $\cL_{1}$ consists of the effective range term, 
proportional to $\rho_{0}$, and 
$\delta\cL_{1}$, which is present to subtract a divergence which arises 
in the calculation of the scattering amplitude with this Lagrangian.
Below we will regularize
all integrals with a sharp UV cutoff. We will renormalize the LO and NLO 
expressions below, calculating the coupling constants $\lambda_{0}$, $\rho_{0}$,
and counter term $\delta\lambda_{0}$ of the theory, to reproduce the t-matrix in 
each case.

\subsection{LO amplitude}
\label{sec:lead-order-ampl}
At LO in $r_{s}$, the $\lambda_0$ vertex has to be iterated to all
orders in order to reproduce the non-perturbative properties of the
large scattering length
limit~\cite{Kaplan:1996xu,vanKolck:1998bw}. 
The diagrams contributing to the two-body scattering amplitude form
the integral equation shown in Fig.~\ref{fig:ALO}, that is equivalent
to the Lippmann-Schwinger (LS) equation,
\begin{figure}[t]
\centerline{\includegraphics[width=\columnwidth,angle=0,clip=true]{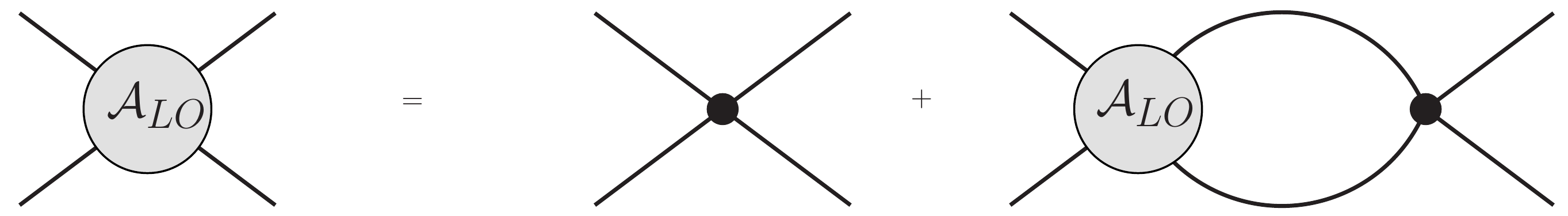}}
\caption{\label{fig:ALO} Scattering amplitude at leading order.}
\label{fig:ALO}
\end{figure}
\begin{equation}
 \label{eq:ALO-EFT}
  i\Alo (E)=-i
  \frac{\lambda_0}{m}-i\frac{\lambda_0}{m} \mathcal{I}_0(E,\Lambda)\, i\Alo (E)~,
\end{equation}
where $E$ denotes total energy of the two-body system. On the mass
shell $E=k^2/m$. The function $\mathcal{I}_0(E,\Lambda)$ in
\eq{ALO-EFT} is given in \eq{I0} in Appendix
\ref{sec:loop_integrals}. It is the loop integral shown in
Fig.~\ref{fig:ALO} and depends on the energy $E$ and the 
ultraviolet (UV) cutoff $\Lambda$ that is imposed on the integral.  

The low-energy
constant $\lambda_0$ is a function of the cutoff and its form is
determined by requiring that \eq{ALO-EFT} reproduces the two-body
t-matrix in \eq{tmatrix-general} in the limit $r_s\to 0$.
We can therefore write $\lambda_0$ explicitly as a function of the
cutoff $\Lambda$ and the scattering length $a$
\begin{align}
\label{eq:lam0}
\lambda_0 &   =\frac{4 \pi a}{1-2 a \Lambda/\pi}~.
\end{align}
\subsection{NLO amplitude}
\label{sec:renorm-at-nlo}
In this subsection we present the renormalization of the EFT including
short-range interactions up to NLO. The renormalization to this order
using dimensional regularization with power divergence subtraction (PDS)
was already discussed in \cite{Kaplan:1996xu}. Since we are using an
explicit momentum space cutoff, we have to introduce an additional
subtraction terms as we will discuss below.

The expansion of the two-body t-matrix in \eq{tmatrix-general} in
$r_s$ can be written as
\begin{equation}
  \label{eq:tmatrix-nlo}
  t(k)=\frac{4\pi}{m}\frac{1}{-\frac{1}{a}-ik}
\left(1-\frac{\frac{r_s}{2} k^2}{-\frac{1}{a}-ik}\right)
\,,\end{equation}
where the first term is LO and second term is its NLO correction.  We
would like to reproduce the second term by calculating corrections to
the two-body amplitude pertubatively due to $\cL_1$.

\begin{figure}[t]
\centerline{\includegraphics[width=\columnwidth,angle=0,clip=true]{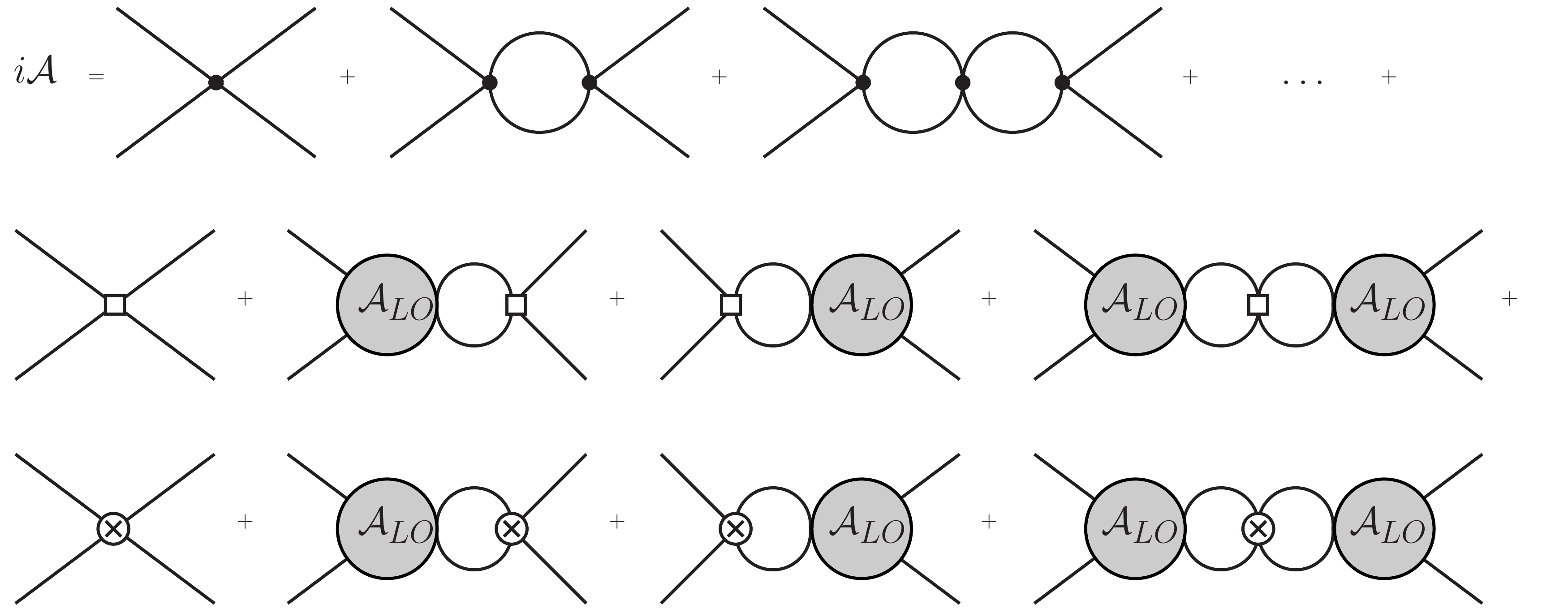}}
\caption{ Scattering amplitude up to
  NLO. The solid dot denotes a $\lambda_{0}$ vertex, the square
  represents the $\rho_{0}$ vertex, and the symbol crossed circles
  represent the counterterm vertex, $\delta\lambda_{0}$.}
\label{fig:Full_Amplitude}
\end{figure}

In \fig{Full_Amplitude}, we show the scattering amplitude up to
NLO. The second row contains the sum of the diagrams with exactly one
insertion of the $\rho_0$ vertex. The third row shows the diagrams
that contain exactly one insertion of the $\delta\lambda_{0}$ vertex.
The factors $\rho_{0}$ and $\delta\lambda_{0}$ are inserted only once
because, as we will see below, they are proportional to $r_s$, which we
only want one factor of inserted to perform the calculation at NLO in
the effective range.  In addition to these two contributions, we have
to consider a contribution that arises from resumming the $\lambda_0$
vertex as a result of using a finite cutoff regularization scheme,
which introduces a correction on the order of $1/\Lambda$.

The sum of these three contributions is given by
\begin{align}
\label{eq:A-NLO}
i\mathcal{A}_{\rm NLO}=
&-i\Alo ^{2}\frac{mk^{2}}{2\pi^{2}\Lambda}
-i2\rho_0\frac{m \Alo ^{2}}{\lambda_0}
\left(\frac{m k^2}{\lambda_0}-\frac{m\Lambda^{3}}{6\pi^{2}}\right)
\nn\\
&-i\frac{\delta\lambda_0}{m}\left( \frac{m \Alo }{\lambda_0}\right)^{2}.
\end{align}
Each term gives the contribution from each of the three rows in
\fig{Full_Amplitude} at NLO. In the first term, we keep only the
NLO contribution proportional to $\frac{k^{2}}{\Lambda}$ after summing
over all diagrams in the first row of the figure.  The second
contribution is the sum of diagrams for the effective range vertex,
and it contains a $\Lambda^{3}$ divergence.
Requiring that the third term, proportional to $\delta\lambda_{0}$, subtracts  
the $\cO(\Lambda^{3})$ divergence and then comparing \eq{A-NLO} to the NLO 
term in 
\eq{tmatrix-nlo}, $\rho_{0}$ and $\delta\lambda_{0}$ are determined to be  
\begin{align}\label{eq:rho0}
\rho_0  & =  \frac{\lambda_0^2}{16\pi m}\left(r_{s}-\frac{4}{\pi\Lambda}\right)
\,,\\ \label{eq:dlam0}
 \delta \lambda_0 &= \frac{(\lambda_0 \Lambda)^{3}}{48\pi^{3} }\left(r_{s}-\frac{4}{\pi\Lambda}\right).
\end{align}

\section{Universal relations at next-to-leading  order}
\label{sec:rels}
In addition to the given field theoretical framework, we use
the OPE to calculate the momentum distribution and find the contact,
$C$, the derivative contact, $D$, and the operator $C'$ that measures
the mean kinetic energy of opposite spin pairs in terms of renormalized field
theoretical operators. Then we express the other relations in terms of
the operators associated with these parameters. However, we leave the
details of this to \sec{OPE} and directly present the results here. We
give the $1/k^6$ correction to the $1/k^4$ tail of momentum
distribution and effective range corrections to the energy relation,
adiabatic relation, pressure relation, and to the virial theorem for a
harmonic potential.

By using the OPE, we find that the momentum distribution of atoms in a
spin state $\sigma$ is given by
\begin{align}
\label{eq:rhoOPE}
 \rho_\sigma (k) \to  \frac{C}{k^4}+\frac{C'+D}{k^6}+ 
 \cO\left(\frac{1}{k^8}\right)
\,,\end{align} 
where $C$ is the well-known contact, which is the asymptote of the
scaled momentum distribution shown in \fig{mom} and a measure of the
sensitivity of the system to the scattering length. We call $D$ the
derivative contact because it is associated with the second derivative
of the contact operator in $\cL_{0}$, and it is a measure of the
sensitivity of the system to the finite effective range, $r_s$. $C'$
is associated with the averaged pair kinetic energy in the system. In
the two-body system $C'=CK^2/2$, where $K$ is the center of mass
momentum.  In the absence of a known value for $C'$, we do not include
it in \fig{mom} below. However, a value could be obtained from the
$1/k^6$ tail of improved QMC simulation data.  
Note that \eq{rhoOPE}
is valid in the limit of zero effective range and that the form
remains the same after taking the effective range correction into
account because the correction is contained in $C$, $C'$, and $D$. The
derivation of \eq{rhoOPE} is given separately in \sec{OPE}, where we
also show that the contact $C$ and derivative contact $D$ are the
expectation values $\langle \int d^{3}\Rvec\,\cO_C \rangle$ and
$\langle \int d^{3}\Rvec\,\cO_D\rangle$, respectively.  $\cO_{C}$ is
proportional to the contact term in $\cL_{0}$, with coupling constant
$\lambda_{0}$, in the Lagrangian, and the operator $\cO_{D}$ is
related to $\cL_{1}$.
Similarly, near $p$- or $d$-wave resonances 
\cite{PhysRevLett.117.019901,dwave}, the
 $1/k^4$ term in the tail of the momentum distribution receives contributions 
 with interpretations similar to $C'$ and $D$.

Next, the energy relation rewrites the sum of kinetic and interaction
energies, each separately sensitive to UV behavior in this field 
theoretical framework, in terms of pieces
which are individually finite.
\begin{align}
\label{eq:Erel}
\langle H \rangle &=\frac{C}{4\pi ma} + r_s\frac{D}{16\pi m}+
\langle T^{\text{(sub)}}\rangle
\,,\end{align}
where $\langle H \rangle$ is the expectation value of the Hamiltonian
for a generic mixture of eigenstates, and $T^{\text{(sub)}}$ is the
subtracted (renormalized) kinetic energy defined here in \eq{Tsub_0}, 
which is calculated using \eq{Tsub} found at the end of \sec{OPE}. 
\begin{align}
\label{eq:Tsub_0}
&\langle T^{\text{(sub)}}\rangle
\nn\\
&=\!\!\sum_\sigma \!\! \int^\infty_0 \!\!\! \frac{dk}{4\pi^2m}
\Big[k^4\rho_{\sigma}(k) -  C
-\frac{\theta(k-k_0)}{k^2} (C'+D)+\cdots\Big] 
\nn\\& \quad
+\frac{C'+D}{2\pi^2 m k_0} -r_s\frac{C'+3D}{16\pi m}
\,.\end{align}

To arrive at this, the kinetic energy operator $\cT$ defined in
\sec{OPE} was rewritten in terms of the renormalized operator in
\eq{ME15-renorm}.  Then this was replaced by the the momentum
distribution $\rho_{\sigma}(k)$ using the OPE result given in
\eq{OPE-k}.  The lower limit $k>k_0$ was imposed to prevent an
IR divergence, and the second to last term was added to remove sensitivity to
the IR cutoff.  The derivation of \eq{Erel} is given after the
derivation of the momentum distribution at the end of \sec{OPE}. While
the energy relation holds for a generic mixture of states, the
relations that follow in the remainder of this section are for a pure
eigenstate.  In the case of a generic mixture they approximately hold,
and only so long as the off diagonal terms are negligible
as discussed in \cite{2008AnPhy.323.2952T,2008AnPhy.323.2971T,2008AnPhy.323.2987T}. 

The adiabatic relation~\cite{2008AnPhy.323.2971T} is defined as the change
in the energy of the system due to an adiabatic change of $a$ so that
the energy eigenstate of the system is not disturbed. A similar
adiabatic relation for the effective range $r_s$ can be obtained by
taking the derivative with respect to $r_s$. Taking the derivative
of the energy of the system with respect to $a$ gives 
\begin{align}
  \label{eq:dEnda} \frac{dE}{d a} = \Big\langle
  \frac{d H }{d a} \Big\rangle= \frac{C}{4\pi m a^2}
  \,,
\end{align}
where, for the first equality, we used the Feynman-Hellmann theorem.
Then, for the second equality we obtain
$dH/da=\int d^3\Rvec\,\cO_C /(4\pi m a^2)+O(1/\Lambda^2)$ by using
$d\lambda_0/da =\lambda_0^2/(4\pi a^2)$ and
$d\rho_0/da=\frac{\lambda_{0}}{2\pi a^{2}}\rho_{0}$.  
Note that \eq{dEnda} remains the same as it was in the zero-range
limit, $r_s\to 0$, in Ref.~\cite{2008AnPhy.323.2987T}, and the effective
range correction is contained in the expectation value $C$.


Taking the derivative of the energy respect to $r_s$, with $a$ held fixed,
gives
\begin{align} \label{eq:dEndr}
\frac{dE}{dr_s} =\Big\langle \frac{d  H }{d r_s}\Big\rangle
 = \frac{ D}{16\pi m}
\,,\end{align}
where we again used the Feynman-Hellmann theorem in the first equality
and then obtained $dH/dr_s= \cO_D/(16\pi m)+ O(\rho_0)$ by using
$d\rho_0/dr_s= \lambda_0^2/(16\pi m)$ and
$d a/dr_s=d \lambda_0/dr_s = d\Lambda/dr_s=0$.  Any terms proportional
to $\rho_0$ should be dropped because they are higher order than NLO when
in $H$. This means that $D$ is independent of $r_s$ at NLO in $r_s$.
In other words, $dE/dr_s$ is well behaved as $r_s \to 0$, as
discussed in
\cite{2012PhRvA..86a3626W,PhysRevA.84.061602,2012PTEP.2012aA209C}.

Next, we derive the effective range corrections to the pressure
relation \cite{2008AnPhy.323.2987T} and the virial theorem
\cite{2008PhRvA..78b5601W}. For a homogeneous gas, the energy scales
linearly with the volume of the system.  Thus, the energy density is a
function of volume-independent variables. For example, the Helmholtz
free energy density $\cF$ depends only on intensive thermodynamic
quantities such as temperature, $T$, chemical potential, $\mu_\sigma$, and
interaction parameters such as $a$ and $r_s$.  Dimensional analysis
implies the following equality
\begin{align}\label{eq:dimF}
\Bigg[T\frac{\partial }{\partial T}+{\mu_\sigma}\frac{\partial
  }{\partial \mu_\sigma}
-\frac{a}{2}\frac{\partial }{\partial a}-\frac{r_s}{2}\frac{\partial }{\partial r_s} 
\Bigg] \cF=\frac52 \cF
\,,\end{align}
where the LHS is the sum of the logarithmic derivative with
respect to each individual parameter of the system multiplied by its
dimension, and it reduces to the free energy density multiplied by its
energy dimension, 5/2. By using the definition $\cF=\cE -T s$, where
$s$ is the entropy density, and the relation
$\cF=-P +\mu_\sigma n_\sigma$, where $\mu_{\sigma}$ is the chemical
potential, for a homogeneous system and eliminating $T s$ and
$\mu_\sigma n_\sigma$ in favor of $\cE$ and $P$, we obtain the
pressure relation as
\begin{align}\label{eq:pressure}
P =\frac23 \cE +\frac{\cC}{12\pi m a}+r_s \frac{\cD}{48\pi m}
\,,\end{align}
where $\cE$, $\cC$, and $\cD$ are $E$, $C$, and $D$ divided by volume of the system.

For a gas trapped in a harmonic potential, $V(\Rvec)$, with frequency
$\omega$, we see from dimensional analysis that the energy obeys 
the relation
\begin{align}\label{eq:dimE}
\Bigg[\sum_i \omega_i\frac{\partial }{\partial \omega_i}-\frac{a}{2}\frac{\partial }{\partial a}-\frac{r_s}{2}\frac{\partial }{\partial r_s} \Bigg] E=E
\,.\end{align}
By using
$\sum_i \omega_i \,\partial V(\Rvec)/\partial\omega_i = 2V(\Rvec)$ and
\eqs{dEnda}{dEndr} we obtain the virial theorem for a trapped atomic gas:
\begin{align}\label{eq:virial}
E =2 V-\frac{C}{8\pi m a}-r_s \frac{D}{32\pi m}
\,,\end{align}
where $V$ is an average of the harmonic potential for the system.

The subleading $1/k^6$ tail in the OPE expansion of the momentum
distribution and its relation to energy in \eqs{rhoOPE}{dEndr} were
first derived in \cite{2012PhRvA..86a3626W} by analyzing the behavior
of the many-body wave function for systems of two-component fermions
with a large two-body scattering length $a$.  We reproduce the same
results by using the OPE for two-body states. 

\section{Ground state results for a homogeneous gas}
\label{sec:result}
In this section we extract the numerical value of the derivative contact
$D$ from the recent QMC calculations and compare it to the value of
the contact, $C$, in the zero range limit.  We also plot the momentum 
distribution in comparison to QMC simulation data.
\begin{figure}[t]
\centerline{\includegraphics[width=\columnwidth,angle=0,clip=true]{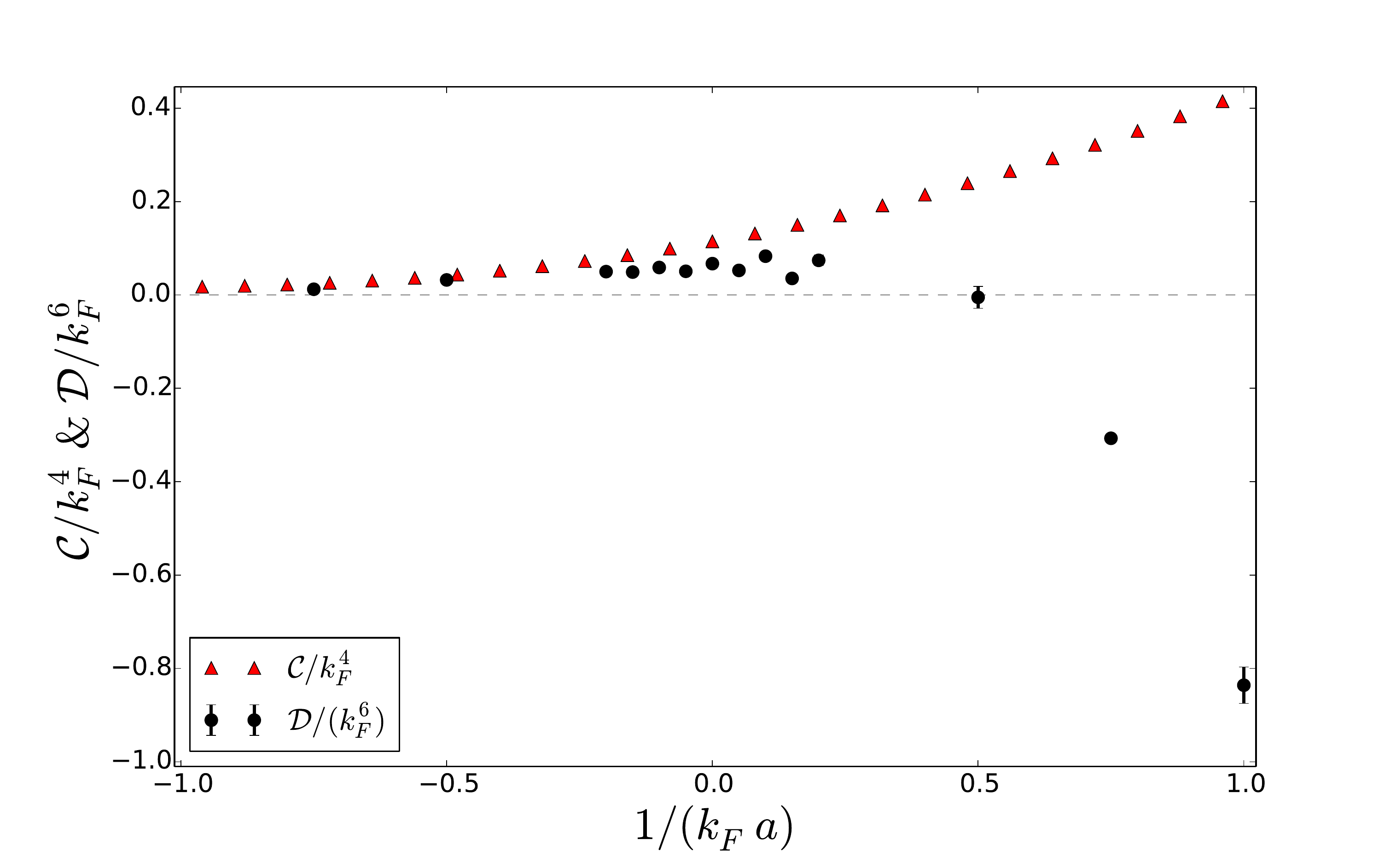}}
\caption{\label{fig:CD} 
(color online) Dimensionless contact density $\cC/k_F^4$, represented by the triangles on
the graph, and the derivative contact density $\cD/k_F^6$ 
as function of $1/(k_F a)$ at zero temperature from QMC simulation in 
\cite{PhysRevA.84.061602,2012PTEP.2012aA209C}.}
\end{figure}

Near the unitary limit, the energy density of a balanced homogeneous
Fermi gas in its ground state can be expressed as
\begin{align}\label{eq:cE}
\cE=\Big( \xi  -\frac{\zeta}{k_F a} + S\, k_F r_s +\cdots 
\Big)\cE_F
\,,\end{align}
where $\cE_F= \tfrac{1}{10\pi^2} k_F^5/m$ is the Fermi energy density,
$k_F= (3\pi^2 n)^{1/3}$ is the Fermi momentum, and $n$ is the total number
 density.
$\xi$ is the Bertsch parameter; $\zeta$ and $S$ are slope constants with
respect to the $1/(k_F a)$ and $k_F r_s$ axes, respectively, in the unitary 
limit. Away from unitarity they are not constant, but the energy density still
contains an effective range correction proportional to $k_F r_s$. Also, the
slope $S$ becomes a function of $k_F a$.

The contact density $\cC$ of the Fermi gas can be obtained in various limits 
using the expressions for the energy in those limits, such as that in the 
unitary limit given by \eq{cE}, and using \eq{dEnda} to arrive at 
\eqss{CBCS}{CUNITARY}{CBEC}.
\begin{align}
\label{eq:CBCS}
\cC/k_F^4 
&\to \frac{4}{9\pi^2} (k_F a)^2\,,  & a\to& 0^-  \text{  (BCS limit)}
\\
 &\label{eq:CUNITARY}\to \frac{2\zeta}{5\pi} \,,	& a\to& \pm \infty \text{  (unitary limit)}
\\
 &\label{eq:CBEC}\to \frac{4}{3\pi} (k_F a)^{-1}\,,      & a\to& 0^+ \text{  (BEC limit)} 
\end{align}
where $\zeta$ is the constant in \eq{cE} determined from experiment or
calculated by various theoretical methods.  The energy density in the BCS and 
BEC limits are given in Ref. \cite{Braaten:2010if}. The dimensionless contact
density $\cC/k_F^4$ is parametrically suppressed by $(k_F a)^{2}$ in the
BCS limit and is parametrically enhanced by $1/(k_F a)$ in the BEC limit,
and it increases as one goes through unitarity from the BCS to the BEC limit.

\begin{figure}[t]
\centerline{\includegraphics[width=\columnwidth,angle=0,clip=true]{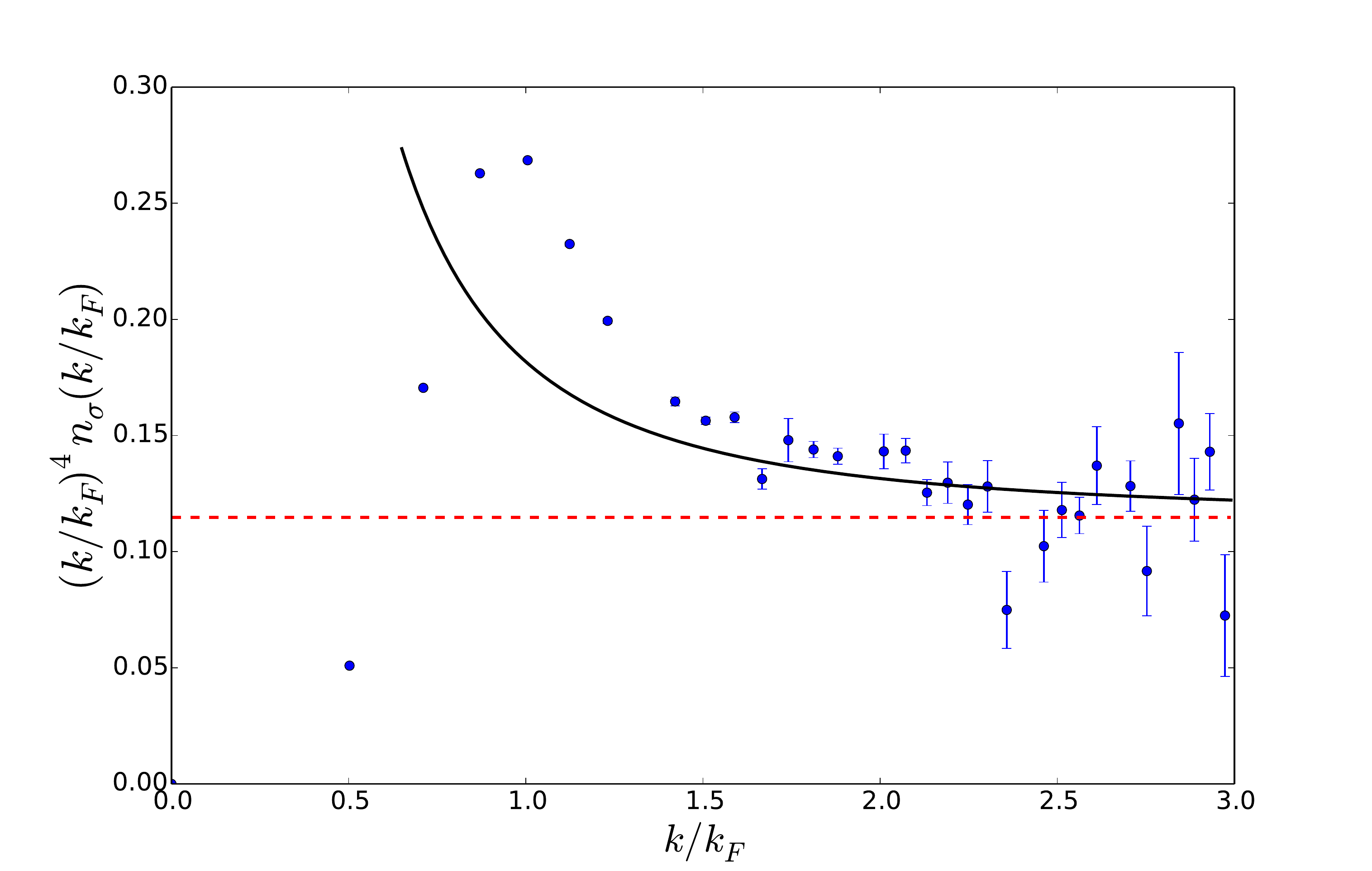}}
\caption{\label{fig:mom} (color online) Scaled momentum distribution 
  near unitarity,
  $(k/k_{F})^{4}n_{\sigma}(k/k_{F})$, where $\sigma$ 
  indicates either of two spin states, as a function of $k/k_F$ in comparison
  to QMC result \cite{2012PTEP.2012aA209C}.  The dashed horizontal line is
  $\cC/k_F^4$. The solid line is
  $\cC/k_{F}^{4}+\cD/(k_{F}^{4}k^{2})$. We use the values
  near unitarity of $\cC/k_{F}^{4}=0.115$ and $\cD/k_{F}^{6}=0.061$.}
\end{figure}

The contact has been determined from many observables by various
experimental groups. The contact across the BCS-BEC crossover was
first determined from photo association
\cite{2005PhRvL..95b0404P,2009EPJB...68..401W}. Precise values in the
unitary limit were obtained: $\zeta = 0.93(5)$ from a
thermodynamic measurement
\cite{2010Sci...328..729N} and $\zeta = 0.91(4)$ from the static
structure factor \cite{2010PhRvL.105g0402K}.  The temperature
dependence was determined from the structure factor using Bragg
spectroscopy \cite{2008PhRvL.101y0403V,2010PhRvL.105g0402K} and also
from RF spectroscopy \cite{2012PhRvL.109v0402S}.  Various universal
relations have been verified by testing the consistency of numerical
values of the contact determined from different observables and
properties of the system such as the momentum distribution, the RF
line shape, photoemission spectra, the adiabatic theorem, and the
virial theorem \cite{2010PhRvL.104w5301S}.

The contact has been calculated not only using QMC
\cite{PhysRevLett.97.100405,0295-5075-75-5-695,2011PhRvL.106t5302D,2013PhRvL.110e5305H,PhysRevA.81.051604,PhysRevC.79.054003},
but also with other
methods\cite{PhysRevA.82.021605,PhysRevA.75.043620,2013arXiv1303.6245V}.
Currently, the most accurate theoretical value is $\zeta=0.901(3)$ given in
\cite{2011PhRvA..83d1601G}. 
Near unitarity this gives a value of $\cC/k_{F}^{4}\approx0.115$.

Additionally, the slope $S$ in \eq{cE} has been calculated in
\cite{PhysRevA.84.061602,2012PTEP.2012aA209C}, and the density of the
derivative contact for the ground state is
\begin{equation}
\label{eq:S}
\cD=\frac{8 k_F^6}{5\pi} \,S(k_F a).
\end{equation}

Some of the asymptotic behavior of the derivative contact density is given by
\begin{align}
\label{eq:Dasymp}
\cD/k_F^6  
	  &\to \frac{8\,S}{5\pi} \,,		 & a\to& \pm \infty \text{  (unitary limit)} 	
    \nn\\ &\to -0.9(1)\, (k_F a)^{-3}\,,      & a\to& 0^+ \text{  (BEC limit)} 
\end{align}
where $S=0.12(1)$ was obtained in \cite{2012PTEP.2012aA209C}.  This
gives a value of $\cD/k^6=0.061$ in the unitary limit.  Equation
\eqref{eq:Dasymp} can be derived using \eq{cE} and the relation
between $\cD$ and $dE/dr_s$ found in \eq{dEndr}. We calculated the
result, namely $\cD/k_{F}^{6}\to-0.9(1)(k_{F}a)^{-3}$, in the BEC
limit, $1/k_{F}a>0$, by fitting $\cD$ to the QMC data for $S$
\cite{2012PTEP.2012aA209C} shown in \fig{CD} (note that this figure
 contains the data rescaled with the prefactor given in \eq{S}).  The dimer
  binding energy
$E_\text{dimer}=1/(ma^2)[ 1+r_s/a+\cdots]$ gives the derivative
contact for the dimer as $D_\text{dimer}=16\pi/a^3$, which is
consistent with the power law in the BEC limit. A power law in the BCS
limit is not known.

Figure \ref{fig:CD} shows results for the scaled contact density
$\cC/k_F^4$ and derivative contact density $\cD/k_F^6$ as a function
of $1/(k_F a)$.  In the unitary limit $\cC$ is about twice as large as
$\cD$.  In the BEC limit the magnitude of $\cD$ increases faster than
that of $\cC$, and this follows from the fact that the effective range
 correction is more important in this limit as the approximation of universal
 physics worsens.
\fig{mom} shows a scaled momentum distribution in unitary limit.  The
OPE results in \eq{rhoOPE} describe the QMC data
\cite{2012PTEP.2012aA209C} well for large $k>1.5 k_F$. However, we
note again that we did not include the unknown $C'$ contribution in
this analysis.

\section{Operator Product Expansion and Related Calculations}
\label{sec:OPE}
In what follows we describe the expansion of the nonlocal
operator for the momentum distribution in the large $k$, or short-distance, limit, that we use in the derivation of the above universal
relations. With the OPE we derive expressions for the contact and
derivative contact in terms of field theoretical operators. We then
express the Hamiltonian in terms of those operators.
The OPE, invented independently by Ken Wilson \cite{Wilson:1969zs},
Leo Kadanoff \cite{Kadanoff:1969zz}, and Alexander Polyakov
\cite{Polyakov:1970} in 1969, is a short-distance expansion of a
nonlocal operator into a series of local operators multiplied with
short-distance coefficients, or Wilson coefficients, that are
functions of the relative separation ${\bf r}$:
\begin{equation}
  \label{eq:OPE-formal}
  \mathcal{O}_A\Big(\Rvec-\frac{\rvec}{2}\Big)\mathcal{O}_B\Big(\Rvec+\frac{\rvec}{2}\Big)
=\sum_n W_n(\rvec)\cO_n(\Rvec)
\,,
\end{equation}
where $W_n$ is the Wilson coefficient of the local operator $\cO_n$.
In this paper we consider the nonlocal one-body density operator
$\psisdagg(\Rvec-\rvec/2) \psis(\Rvec+\rvec/2)$, which is a coordinate
space representation of the momentum distribution and gives the
momentum space distribution $\rho_{\sigma}(k)$ after a Fourier
transform.

The operators on the RHS in \eq{OPE-formal} can be constructed from
the fields of the EFT and their gradients. The field $\psis$ has
dimension $\Delta = 3/2$ in our EFT expansion, the Galilean invariant
derivative
\begin{equation}
\label{eq:partial}
\arrowpartial{i}=\rarrowpartial{i}-\larrowpartial{i}
\end{equation}
has $\Delta=1$, and the time derivative $\partial_t$ has $\Delta=2$.  Here,
we list the relevant operators up to dimension $\Delta= 6$.
\begin{equation}
\label{eq:operators}
\begin{array}{cclcl}
\Delta &\qquad& \cO_{1, \Delta} &\qquad & \cO_{2, \Delta} \\
3	 &&	\psisdagg\psis &&\mbox{}\\
4	 &&\psisdagg\arrowpartial{i}\psis &&\Oc \\  
5	 && \psisdagg \arrowpartials{i}{j}\psis && \psiadagg\psibdagg\psib\arrowpartial{i}\psia+h.c.\\
6	 && \psisdagg \arrowpartialss{i}{j}{k}\psis &&\psiadagg\psibdagg\psib \arrowpartials{i}{j}\psia+h.c. \\
\end{array}
\end{equation}

A unit operator is omitted since the momentum distribution for the vacuum is 
zero while the unit operator vacuum expectation value is nonzero, 
and the operator with time derivative
$\partial_t$ is excluded because we can always eliminate it in favor
of momentum-dependent operators by using the equations of motion
obtained from \eqs{L0}{L1} \cite{Goldberger:2010fr}.  In the first column of the operator list
the number in each row represents the dimension of operators in that
row.  The second and the third column list the one-body and two-body
operators as indicated in the subscript of $\cO_{1,\Delta}$ and
$\cO_{2,\Delta}$.  It is known \cite{Braaten:2008uh} that the
dimension of $\Oc$ is lowered down to $\Delta=4$ from $\Delta=6$ by
the strong interaction, and we find that this is true for the operator
$\psiadagg\psibdagg\psib \arrowpartials{i}{j}\psia$ so that it has
$\Delta=6$ rather than $8$.

Rewriting \eq{OPE-formal} for our problem we have
\begin{align} \label{eq:OPE}
	&\psisdagg(\Rvec-\rvec/2)\psis(\Rvec+\rvec/2)
	\nn\\&
	=\sum_{\Delta}\left(W_{1,\Delta}(\rvec)\, 
	\cO_{1,\Delta}(\Rvec)+W_{2,\Delta}(\rvec)\, 
	\cO_{2,\Delta}(\Rvec)+\cdots\right)
\,,\end{align}
where the first index $n$ of each $W_{n,\Delta}$ indicates whether the
coefficient is for a one- or two-body operator, and the second index
$\Delta$ gives the scaling dimension of that operator.  We determine
the Wilson coefficients of one- and two-body operators up to $\Delta=6$.

Before we calculate the matrix elements of the operators shown above,
we discuss briefly the off-shell scattering amplitude, whose
knowledge is required for the derivation of our Wilson coefficients.
In the renormalization of the EFT in \sec{effect-field-theory}, we
considered the on-shell amplitude depending on the relative momentum
$k$, with $k^2=m E$, but we use the off-shell amplitude during
operator matching since its explicit momentum dependence contributes
to the loop integral results in our calculations. In general, the
off-shell amplitude with incoming momenta $(E/2,\pm \pvec)$ and 
outgoing momenta $(E/2,\pm \kvec)$ should be a function of three
variables: $E$, $\pvec$, and $\kvec$. However it is known that the LO
amplitude, $\cA_\text{LO}(E)$, depends solely on the total energy and not
on external momenta.  This simplifies the calculation of a loop
diagram involving $\cA_\text{LO}(E)$, such as the last graph in
\fig{ALO}, by factorizing it into a product of the amplitude and the
loop integral as shown in the last term of \eq{ALO-EFT}. However, this
is not true at NLO.  Expanding the off-shell amplitude in powers of
$1/\Lambda$ we obtain
\begin{align}
	\label{eq:Amplitude-Off-Shell}
&\cA_\text{LO}-\Alo ^{2}\frac{m(mE)}{2\pi^{2}\Lambda}
-\Alo ^{2}\frac{ 2m^2 \rho_0 mE}{\lambda_{0}^{2} }
\nn\\
&\qquad\qquad +\Alo \frac{m\rho_0}{\lambda_0}(\pvec^{2}+\kvec^{2}-2mE)+\cdots
\,,
\end{align}
where the first term, the LO amplitude, is proportional to $1/E$ and the
next two terms are NLO in $1/\Lambda$ and reduce to \eq{A-NLO} in the
on-shell limit. The last term, proportional to
$\pvec^2/\Lambda^2$, and terms beyond this are power suppressed.
Here, we count the size of parameters as follows:
$a^{-1},\pvec,\kvec,\sqrt{mE}$ are of the same size and much smaller
than $\Lambda$, while $\lambda_0\sim 1/\Lambda$ and
$\rho_0\sim 1/\Lambda^3$.  The terms of order $1/\Lambda^2$ and higher
can be dropped when the final goal is to compute the amplitude at NLO,
and $\cA$ then depends only on $E$, becoming $\Anlo(E)$.  But, in the
calculation of matrix elements, the term proportional to
$\pvec^2+\kvec^2$ contributes factors of the loop momentum in the loop
diagrams involving the amplitude and can lead to a UV
divergence. Therefore, terms of $\cO(1/\Lambda^2)$ can only be dropped
at the end of the calculation.

At the end of calculations we keep terms only up to NLO in
$(1/\Lambda)$ and renormalize the operators either by 
multiplying by a renormalization factor or by adding different
operators as counterterms to subtract the divergences and
$\Lambda$-dependence.  We rewrite the full off-shell amplitude $\cA(E,\pvec,\kvec)$
in terms of an amplitude $\Alam(E)$ that contains only diagrams that
contain the $\lambda_0$-coupling constant and amplitudes $\Arho(E)$
and $\Arhop(E)$ that contain one insertion of the $\rho_0$-coupling
constant but scale as $1/\Lambda$ and $1/\Lambda^2$, respectively.  In
these amplitudes, all power suppressed terms
$(\sqrt{mE}/\Lambda)^n$ in the loop integral $\cI_0(E,\Lambda)$ are
retained. The amplitude is given as
\begin{align} \label{eq:Alamrho}
\cA(E,\pvec,\kvec) &=\cA(E)+(\pvec^2+\kvec^2-2mE)\Arhop(E) 
\,,\nn\\
\cA(E)& =\Alam(E) +\Arho(E)  
\,,\nn\\
\Alam &= -\frac{1}{m/\lambda_0 +i \cI_0(E,\Lambda)}
\,,\nn\\
\Arho &= -\frac{2\rho_0 m^2}{\lambda_0^2} mE \,\Alam^2 
\,,\nn\\
\Arhop &= \frac{\rho_0 m}{\lambda_0} \,\Alam 
\,,\end{align}
where $\cI_0(E,\Lambda)$ is given in the Appendix in \eq{I0}. 
\begin{figure}[t]
\centerline{\includegraphics[width=\columnwidth,angle=0,clip=true]{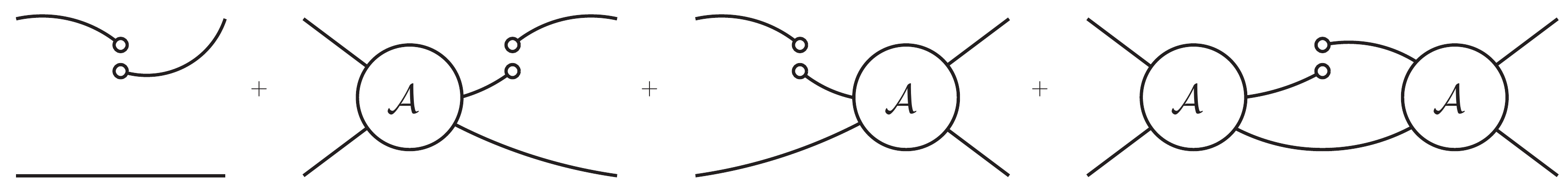}}
\caption{\label{fig:rho_k} 
Diagrams for the operator
$\langle\psi^{\dagger}_{\sigma}(\Rvec-\rvec/2)\psi_{\sigma}(\Rvec+\rvec/2)\rangle$
for the 2 atom scattering state.
The empty dots imply locations $\pm \rvec/2$ where an atom is
annihilated and created at an equal time.}
\label{fig:non-local_ALO}
\end{figure}
Also, the
amplitude $\Alam$ still satisfies the LS equation in \eq{ALO-EFT} when
$\Alo$ is replaced by $\Alam$.
\begin{equation}
\label{eq:LS-lam}
\Alam=-\lambda_0/m ( 1+ \cI_0 \, i \Alam)
\,.
\end{equation}
 
Figure \ref{fig:rho_k} shows diagrams contributing to the matrix
element for the two-atom scattering state with incoming momenta
$(p_0,\pm \pvec)$, with outgoing momenta $(k_0,\pm \kvec)$, and with
total energy $E=2 p_0=2 k_0$. We generalize to a system with non-zero
center-of-mass momentum later in this section. The matrix element of the
nonlocal operator is given by
\begin{align}\label{eq:cal-nonlocal}
 & \langle \psi^\dagger_\sigma(\Rvec-\tfrac{\rvec}{2})\psi_\sigma(\Rvec+\tfrac{\rvec}{2}) \rangle
 \nn\\ &= \delta_{\pvec\kvec}e^{i \pvec \cdot \bm{r} }
     +\Bigg[\frac{i e^{i \pvec \cdot \bm{r}}}{p_0-\frac{\pvec^2}{2m}} i\cA(E)
     +(p\to k)\Bigg]
     -\cI_{\rho,0} \, \cA^2(E)
     \nn\\
   &\quad -2 \big[(\tfrac{\pvec^2+\kvec^2}{2} -2mE)\cI_{\rho,0}+\cI_{\rho,2}  \big]\Alam \Arhop 
     \,,
\end{align}
where we use the shorthand notation
$\delta_{\pvec\kvec}= (2\pi)^3 \delta^{(3)}(\pvec-\kvec)$, and the
one-loop integral $\cI_{\rho,2n}$ is given in the Appendix in \eq{Irho}.

By inserting $\cI_{\rho,0}$ and $\cI_{\rho,2}$ from \eq{Irho} into
\eq{cal-nonlocal} and expanding in powers of $r$ up to $r^3$, we
obtain
\begin{widetext}
\begin{align}	\label{eq:non-local-leading-order}
\langle\psi^{\dagger}_{\sigma}(\Rvec-\tfrac{\rvec}{2})\psi_{\sigma}(\Rvec+\tfrac{\rvec}{2})\rangle
&=\delta_{\pvec \kvec}
	-\Bigg[ \frac{\cA(E)}{p_{0}-\tfrac{\pvec^2}{2m}} +(p\to k) \Bigg]
	+\frac{i\cA^{2}(E)m^{2}}{8\pi\sqrt{mE}}
\nn \\ 
&\quad +ir_{i}p_{i}\delta_{\pvec \kvec}
	-i r_i\, \Bigg[ \frac{p_i\, \cA(E)}{p_{0}-\tfrac{\pvec^2}{2m}} +(p\to k) \Bigg]
	-r\, \frac{\cA^{2}(E)m^{2}}{8\pi}
\nn \\
&\quad -r_{i}r_{j}\frac{p_{i}p_{j}}{2}\delta_{\pvec \kvec}
	+\frac{r_i r_j}{2}\Bigg[ \frac{p_i p_j \cA(E)}{p_{0}-\tfrac{\pvec^2}{2m}} +(p\to k) \Bigg]
		-i r^2\,\frac{\cA^{2}(E)m^{2} \sqrt{mE}}{16 \pi}
\nn \\&\quad 
	-ir_{i}r_{j}r_{k}\frac{p_{i}p_{j}p_{k}}{6}\delta_{\pvec \kvec}
		+i \frac{r_i r_j r_k}{6}\Bigg[ \frac{p_i p_j p_k \cA(E)}{p_{0}-\tfrac{\pvec^2}{2m}} +(p\to k) \Bigg] 
		+r^3\,\frac{\cA^{2}(E)m^{3} E}{48 \pi}
\nn \\&\quad + \cO(r^4)
\,,\end{align}
\end{widetext}
where the Einstein summation convention for like indices is
assumed, and we have dropped terms $\cO\left(1/\Lambda^2\right)$. 
In this equation there are terms that have $\sqrt{mE}$ dependence, which
we distinguish from $|\pvec|$ because we keep the atoms off the mass
shell in our calculations, {\it i.e.}  $E\neq\frac{\pvec^{2}}{m}$.  Each
term in \eq{non-local-leading-order} will be matched to the matrix
elements of the relevant local operators, and by this matching process
the Wilson coefficients of the local operators will be determined.
$\cA^2(E)$ is the renormalized square amplitude and includes the
correction due to the finite effective range. But,
\eq{non-local-leading-order} is still valid in the zero range limit,
in which we replace $\cA(E)$ by $\cA_{\text{LO}}$. With the $r^3$ term
still present, this confirms that the $1/k^6$ correction to the tail
of the momentum distribution is still present even when $r_s \to 0$.


Next, we calculate the matrix elements of local operators for the
two-body scattering state. As in the calculation for the nonlocal
operator, we do not drop power-suppressed terms in the amplitude
\eq{Alamrho} during intermediate steps, and we keep terms up to NLO in
the EFT expansion after the renormalization.

\subsection{One-body local operators}
In this subsection, we calculate matrix elements for the one-body
operators of various scaling dimensions $\cO_{1,\Delta}$ in \eq{operators}. 
The Feynman diagrams
in \fig{One_body} show all contributions for the two-atom scattering state.
The first diagram contributes even in the absence of two-body interactions.

\begin{figure}[t]
\centerline{\includegraphics[width=\columnwidth,angle=0,clip=true]{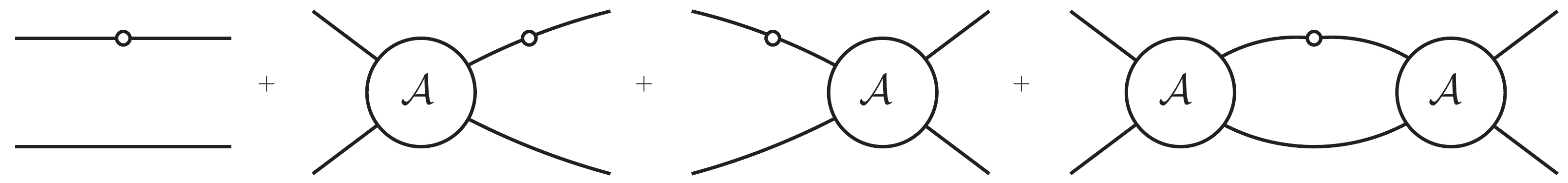}}
\caption{\label{fig:One_body} Diagrams for one-body operator. Empty
  dots imply an insertion of the one-body operator given in
  \eq{operators} and $\cA(E,\pvec,\kvec)$ represents the off-shell
  amplitude in \eq{Alamrho}.}
\label{fig:One_body}
\end{figure}

Since \fig{One_body} is the same for any one-body operator,  
its matrix element can be written as:
\begin{align} \label{eq:ME1d}
&\langle \cO_{1,\Delta} \rangle  =
v_{1,\Delta}(\pvec)\, \delta_{\pvec \kvec}-\Bigg[ \frac{v_{1,\Delta}(\pvec)\,\,\cA(E)}{p_0-\pvec^2/(2m)}+ (p\to k)\Bigg] 
\nn\\&
-\IaD{0}\, \cA^2(E)
-2 \big[(\tfrac{\pvec^2+\kvec^2}{2} -2mE)\IaD{0}+\IaD{2}  \big]\Alam\Arhop 
\,,\end{align}
where $v_{1,\Delta}(\pvec)$ is the vertex factor for the operator
$\cO_{1,\Delta}$. The vertex factors up to dimension 6 are given in \tab{v1d}
in the Appendix.  The loop integrals $\cI_{2n}^{(1,\Delta)}$ are also found
in the Appendix in \eq{I1d}.

The matrix element of 
$\cO_{1,3}=\psisdagg \psis$ for a two-body state
and ingoing momentum $\pvec$ and outgoing momentum $\kvec$ is
\begin{align} \label{eq:ME13}
\langle \cO_{1,3} \rangle  =
&\delta_{\pvec \kvec}-\Bigg[ \frac{\cA(E)}{p_0-\pvec^2/(2m)}+ (p\to k)\Bigg] 
\nn\\
&+\frac{ i m^2  \cA^2(E)}{8\pi \sqrt{mE}} \Big( 1+i r_s \sqrt{mE} \Big)
+O(\Lambda^{-2})
\,.\end{align}
Note that $r_{s}\cA^{2}(E)=r_{s}\cA_{\lambda}^{2}$ up to NLO in $r_{s}$.

Except for the term proportional to $r_s$, \eq{ME13} is matched to the second
line of \eq{non-local-leading-order}, and the Wilson coefficient is

\be \label{eq:W13}
	W_{1,3}=1
\,. \ee 
The $r_s$ term will be subtracted later by including an appropriate 
term in the Wilson coefficient of the 2-body operator in \eq{ME24-ren}.

Next, the matrix element of $\cO_{1,4}=\psisdagg \arrowpartial{i} \psis$ is given by
\begin{align} \label{eq:ME14} \langle \cO_{1,4} \rangle &=
  2ip_i\delta_{\pvec \kvec} -\Bigg[
  \frac{2ip_i\cA(E)}{p_0-\pvec^2/(2m)}+ (p\to k)\Bigg] \,.
\end{align}
By comparing this to terms containing one power of $p_i$ in
\eq{non-local-leading-order}, it is clear that the Wilson coefficient
corresponding to $\cO_{1,4}$ is given by
\begin{equation}
\label{eq:W14}
	W_{1,4}=\frac{1}{2}r_{i}~.
\end{equation}

The next operator is $\cO_{1,5}=\psisdagg \arrowpartials{i}{j} \psis$ and its matrix element is given by
\begin{align} \label{eq:ME15}
\langle \cO_{1,5} \rangle  
=&-4p_i p_j\,\delta_{\pvec \kvec}
+\Bigg[ \frac{4p_i p_j\cA(E)}{p_0-\pvec^2/(2m)}+ (p\to k)\Bigg] 
\nn\\
&-\frac{4\delta_{ij} }{3}\Big[i\frac{d\, \cI_2}{dE} \Big]\cA^2(E)
-\frac{4m\delta_{ij} }{3}
\Bigg[\frac{m\Lambda^3}{3\pi^2}
\nn\\
&+ \Big(\tfrac{\pvec^2+\kvec^2}{m}-2E\Big)i\frac{d\, \cI_{2}}{dE}+2imE\cI_{0}\Bigg]\Alam\Arhop 
\,.\end{align}
Note that this matrix element is linearly divergent, with
$\Lambda^3\Arhop\sim\Lambda$, and we will renormalize this divergence
and other explicit $\Lambda$-dependence by adding the two-body
operators $\Oc$ and $\Od$ with appropriate factors. The operator
$\psisdagg \arrowpartials{i}{j} \psis$ with $i,j$ contracted becomes
the kinetic term in the Hamiltonian and the cutoff dependence implies
that the kinetic energy is sensitive to the short-distance region of a
fundamental potential smaller than the length scale
$a\sim 1/\sqrt{mE}$ beyond which our effective theory loses predictive
power. Through this renormalization procedure, we find
combinations of operators that are insensitive to the short-distance
behavior.

By using the renormalized two-body operators found in Subsection 2 below in
\eqs{ME24-ren}{ME26-ren}, we obtain the result for the renormalized
operator $\cO_{1,5}$
\begin{align}\label{eq:ME15-renorm}
&\langle \cO_{1,5}^\text{(ren)} \rangle 
\nn\\
&=\Bigg\langle 
\cO_{1,5}
+\frac{2\delta_{ij}}{3\pi^2}\Bigg[ \Lambda\big(1+\tfrac{2 z}{3}\big)\, \cO_{2,4}^\text{(ren)}
+\frac{1-3z}{4\Lambda} \cO_{2,6}^\text{(ren)} \Bigg]
\Bigg\rangle 
\nn\\
&=-4p_i p_j \delta_{\pvec \kvec}
+\Bigg[ \frac{4p_i p_j \cA(E)}{p_0-\pvec^2/(2m)}+ (p\to k)\Bigg] 
\nn\\ 
&\quad	-i\delta_{ij}\frac{\sqrt{mE}}{2\pi} m^2\cA^2(E)+O(\Lambda^{-2})
\,.\end{align}
Here $z=\tfrac{m\rho_0\Lambda^2}{\lambda_0}$ and the superscript
$^\text{(ren)}$ indicates a renormalized operator, which has its UV
divergence content properly regularized and renormalized.  By comparing
\eq{ME15-renorm} to the $O(r^2)$ terms in \eq{non-local-leading-order}, we
find this operator's Wilson coefficient to be
\begin{equation}
\label{eq:W15}
W_{1,5}=\frac{1}{8}r_{i}r_{j}. 
\end{equation}
\begin{center}
\begin{table}[t]
\begin{tabular*}{\columnwidth}{@{\extracolsep{\fill}} c c c }
\hline
$\Delta$ & $\cO_{1,\Delta}$ & $W_{1,\Delta}$ \\
\hline
3	&	$\psisdagg\psis$& $1$\\

4	
       &$\psisdagg\arrowpartial{i}\psis$
          & $\frac{1}{2}r_{i}$\\

5	&
	$\psisdagg\arrowpartials{i}{j}\psis^{\text{(ren)}}$	&	$\frac{1}{8}r_{i}r_{j}$\\

6	&	$\psisdagg\arrowpartialss{i}{j}{k}\psis$	&	$\frac{1}{48}r_{i}r_{j}r_{k}$\\

\end{tabular*}
\caption{One-body operators up to scaling dimension 6 and their Wilson
  coefficients $W_{1,\Delta}$, where the 1 corresponds to one-body and
  the $\Delta$ corresponds to the dimensionality of the operator.  }
\label{tab1}
\end{table}
\end{center}


Lastly, the matrix element of
$\cO_{1,6}=\psisdagg \arrowpartialss{i}{j}{k} \psis$ is given by
\begin{align} \label{eq:ME16} \langle \cO_{1,6}\rangle 
&= -i8 p_i p_j
  p_k \delta_{\pvec \kvec} +\Bigg[ \frac{i8p_i p_j p_k \cA(E)
  }{p_0-\pvec^2/(2m)}+ (p\to k)\Bigg] \,.\end{align} 

 No renormalization is necessary for this operator.  Comparing
this to the terms of $O(r^3)$ in \eq{non-local-leading-order}, we get
its Wilson coefficient:
\begin{equation}
\label{eq:W16}
	W_{1,6}=\frac{1}{48}r_{i}r_{j}r_{k}.
\end{equation}



\subsection{Two-Body Local Operators}
\begin{figure}[t]
\centerline{\includegraphics[width=\columnwidth,angle=0,clip=true]{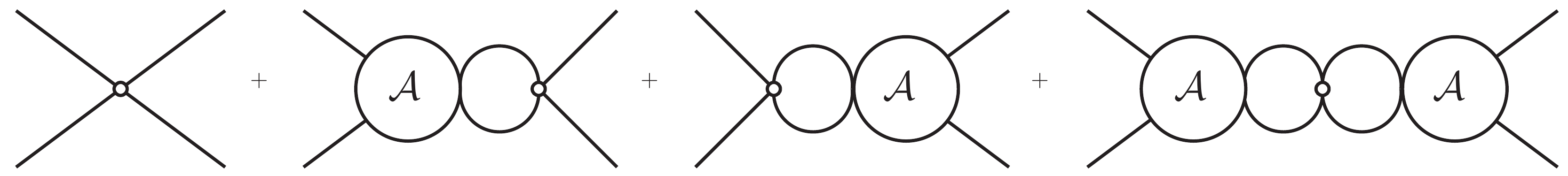}}
\caption{Diagrams for 2-body operators (empty dot).}
\label{fig:Two-body}
\end{figure}

In this subsection we calculate the matrix elements of the two-body
operators $\cO_{2,\Delta}$ for two-atom scattering states and
determine their Wilson coefficients, $W_{2,\Delta}$. Figure
\ref{fig:Two-body} shows the relevant diagrams for the two-body
scattering state. The operators of dimensions $\Delta=4,6$ are those
which are relevant for the momentum distribution and other universal
relations.

Since \fig{Two-body} is the same for any 2-body operators, 
its matrix element can be written generally as
\begin{align} \label{eq:ME2d}
&\langle \cO_{2,\Delta} \rangle  =
(1+i\cI_0 \cA(E))\big[ v_{2,\Delta} (p,k)+2i\IbD{0}\cA(E) \big]
\nn\\&
+i\Arhop \Bigg\{ 2\IbD{2} (1+i\cI_0 \Alam)
\nn\\&
+ \cI_2 \big[ v_{2,\Delta} (p,k)+2i\IbD{0}\Alam \big]
\nn\\ \quad &
+(\pvec^2+\kvec^2-4mE) (1+2 i\cI_0 \Alam)\IbD{0}
\nn\\&
+\Big[ v_{2,\Delta}(p)(\kvec^2-2mE)+v_{2,\Delta}(k)(\pvec^2-2mE) \Big] \cI_0 \Bigg\}
\,,\end{align}
where $v_{2, \Delta}(p,k)=v_{2,\Delta}(p)+v_{2,\Delta}(k)$ is the
vertex factor for the operator $\cO_{2,\Delta}$, and up to scaling
dimension 6 these factors are given in \tab{v2} in the Appendix.  For
convenience, the vertex factors are broken up into terms depending on
a vertex's ingoing and outgoing momenta. Thus, the factors can depend
either on the external momentum or the loop momentum of the
particles. Loop momentum-dependent terms are included in the integrals
defined by \eq{I2d} in the Appendix. The integrals $\cI_0$ and $\cI_2$
are defined in the Appendix by \eq{I2n}.  One important note is that
$\cO_{2,4}$ has the momentum-independent vertex factor
$v_{2,4}(p,k)=1$. This is broken up into $v_{2,4}(p)=1/2$ and
$v_{2,4}(k)=1/2$ simply to follow the prescription given by \eq{ME2d}.
Also, one is not allowed to directly use the LS
equation to simplify $(1+i\cI_{0}\cA)$ in the above equation because
$\cA(E)=\cA_{\lambda}+\cA_{\rho}$ here. The LS equation only
includes $\cA_{\lambda}$.  Additionally, any factors of $\cA$(E)
multiplying $\cA_{\rho'}$ are $\cA_{\lambda}$, rather than
$\cA_{\lambda}+\cA_{\rho}$, since we only want terms up to NLO in
$\rho_{0}$, or $r_s$.


The leading two-body operator $\cO_{2,4}=\Oc$ has been already
calculated in the zero-range model \cite{Braaten:2008uh} and the field
theoretical two-channel model for a narrow Feshbach resonance
\cite{Braaten:2010if}.  Its matrix element is given by
\begin{align} \label{eq:ME24}
&\langle \cO_{2,4} \rangle  
= (1+ i\cI_0 \, \cA(E))^2
\nn\\&
+i\Arhop(1+ i\cI_0 \, 
\Alam)\big[ 2\cI_2+(\pvec^2+\kvec^2-4mE) \cI_0  \big]
\,.\end{align}
With a multiplicative factor $\lambda_0^2$ we obtain a renormalized
operator $\lambda_0^2 \Oc$ at LO in
\cite{Braaten:2008uh,Braaten:2010if}.  Similarly at NLO, it is
renormalized by a multiplicative factor as
\begin{align}\label{eq:ME24-ren}
&\langle
\cO_{2,4}^\text{(ren)}
\rangle 
=\Big\langle 
\lambda_0^2\Big(1+\frac{m\rho_0\Lambda^3}{3\pi^2} \Big)\cO_{2,4} \Big\rangle 
\nn\\
&=m^2 \cA^2(E)+ 2m\lambda_0\Arho-i( \pvec^2+\kvec^2-2mE)m^2\rho_0\cI_0  
\Alam^2 
\nn\\
&=m^2 \cA^2(E)+O(\Lambda^{-2})
\,,\end{align}
where we used the LS equation in \eq{LS-lam} to eliminate
$ 1+\cI_0 \, i \Alam$ in favor of $-m\Alam/\lambda_0$.  As shown in
the last expression, we drop terms $\sim1/\Lambda^2$. Note that the 
operator $\cO_{2,4}^{(\text{ren})}$ is $\cO_{C}$, the contact density
operator.

This can be compared to the third term proportional to $r$ in
\eq{non-local-leading-order}, and this operator's Wilson coefficient 
is given by
\be\label{eq:W24}
	W_{2,4}=- \frac{r}{8\pi}+\frac{r_s}{8\pi}
\,,\ee
where we need the term proportional to $r_s$ here to cancel the term 
proportional to $r_s$
in the one-body result in \eq{ME13}, as discussed before. Note that for 
$\cO_{1,3}$ we did not write $\cO_{1,3}^{(ren)}=\cO_{1,3}+\frac{r_s}{8\pi}
\lambda_{0}^{2}\cO_{2,4}$. This is because the matrix element 
$\langle\cO_{1,3}\rangle$ is finite with no explicit $\Lambda$-dependence, 
so renormalization is not the proper procedure to eliminate the extra term. 
On the other hand, the extra term needs to be 
canceled, so we use the Wilson coefficient of $\cO_{2,4}$ to accomplish
this by adding the term proportional to $r_s$ in \eq{W24}.

The matrix element of 
$\cO_{2,5}=\psiadagg\psibdagg\psib\arrowpartial{i}\psia+h.c. $
is simple because $\Ib{5}{2n}=0$.
\begin{align} \label{eq:ME25}
&\langle \cO_{2,5} \rangle  
= 2i(p_i + k_i)[1+ i\cI_0 \,  \cA(E)]
\nn\\
&-2\Arhop \Big\{ (p_i+k_i) \cI_2+[p_i(\kvec^2-2mE)+k_i(\pvec^2-2mE)] \cI_0 \Big\}
\,.\end{align}
This operator is renormalized by a multiplicative factor as
\begin{align} \label{eq:op25-ren}
\langle \cO_{2,5}^{(\text{ren})} \rangle  
&=\langle\lambda_0\Big(1+\frac{m\rho_0\Lambda^3}{6\pi^2} \Big) \cO_{2,5} \rangle  
\,,\nn\\
&=-2i(p_i + k_i) m\cA(E)+O(\Lambda^{-2})
\,.\end{align}
But, this matrix element is not matched to any term in
\eq{non-local-leading-order} and its Wilson coefficient is thus
$W_{2,5}=0$.

For the operator
$\psiadagg\psibdagg \psib \arrowpartials{i}{j} \psia+h.c.$,
we define $\cO_{2,6}$ as this operator contracted with $\delta_{ij}$
because then it matches to the terms in
the matrix element of the nonlocal operator in
\eq{non-local-leading-order}.
\begin{align} \label{eq:ME26-1}
&\langle \cO_{2,6} \rangle  
=
-4(1+i\cI_0 \cA(E)) (\pvec^2+\kvec^2+2i\cI_2 \cA(E))
\nn\\&
-4i\Arhop \Big\{ 2(1+i\cI_0 \Alam)\cI_4+(\pvec^2+\kvec^2+2i\cI_2 \Alam)\cI_2
\nn\\&\qquad\qquad		 +(\pvec^2+\kvec^2-4mE)(1+2i\cI_0 \Alam) \cI_2
\nn\\&\qquad\qquad +[\pvec^2(\kvec^2-2mE)+\kvec^2(\pvec^2-2mE)] \cI_0
			 \Big\}
\,.\end{align}
\begin{table}[t]
\begin{tabular*}{\columnwidth}{@{\extracolsep{\fill}} c c c }
\hline
$\Delta$ & $\cO_{2,\Delta}$ & $W_{2,\Delta}$ \\
\hline
4	&	$\psiadagg\psibdagg \psib\psia^\text{(ren)} $& $-\frac{r-r_s}{8\pi}$\\
5	&$\psiadagg\psibdagg\psib\arrowpartial{i}\psia^\text{(ren)}+h.c.$
                           &	0\\       
6	&	$\psiadagg\psibdagg\psib\arrowpartialsq\psia^\text{(ren)}+h.c.$	&	$-\frac{r^{3}}{384\pi}$\\
\end{tabular*}
\caption{Two-body operators $\cO_{2,\Delta}$ up to scaling dimension 6
  and their Wilson coefficients.}
\label{tab2}
\end{table}

To renormalize $\cO_{2,6}$, we use momentum-dependent operators
only, and work in the on-shell limit, where $\pvec^2=\kvec^2=mE$,
which allows us to avoid energy dependent operators.
\begin{align}
\label{eq:ME26-ren}
\langle\cO_{2,6}^{(\text{ren})}\rangle 
=&  \Bigg\langle
\lambda_{0}^{2}  \left(1+\tfrac{3}{2}x\right)\cO_{2,6}
\nn\\&
-\left[\frac{4\lambda_{0}^3\Lambda^{3}}{3\pi^{2}}\left(1+2x\right)+\frac{12\lambda_0^2\Lambda^{2}x}{5}\right]\cO_{2,4} 
\Bigg\rangle\\
&=-8mE \,m^2\cA^2(E)+O(\Lambda^{-2})\,,
\end{align}
where $x=\frac{\rho_{0}m\Lambda^{3}}{3\pi^{2}}$.  
The operators for the contact $C$ and derivative contact $D$ in \eq{rhoOPE} are related to 
the renormalized operators as
\begin{align}\label{eq:O24Oc}
\cO_{2,4}^{(\text{ren})}&=\cO_{C}\,,
\\
\label{eq:O26Od}
\cO_{2,6}^{(\text{ren})}&=-4\cO_{D}\,.
\end{align}
This equality is seen when we drop the
factors of $\rho_{0}$ in \eq{ME26-ren} and compare to \eq{OD}, as
there is already a factor of $\rho_{0}$ in the Hamiltonian and we are
only treating the problem to NLO in $r_{s}$.  
Then, \eq{ME26-ren} is matched to the last $r^3$ term in
\eq{non-local-leading-order} with the following Wilson coefficient:
\begin{equation}
\label{eq:W26} 
	W_{2,6}=-\frac{r^{3}}{384\pi}.
\end{equation}
The Wilson coefficients for the two-body operators are compiled 
in Table \ref{tab2}.

\newcommand{\arrowpartialvec}{\overleftrightarrow{\partial}}

Now we generalize our results by considering a two-body system with
a finite center-of-mass momentum $\bm{K}$. The relative momentum of
each particle remains $\pm\bm{p}$.  The single particle momentum,
expressed in terms of center-of-mass and relative momenta, is
$\bm{K}/2\pm\bm{p}$. In addition to this momentum shift by $\bm{K}/2$,
the total energy $E$ is replaced by the Galilean invariant energy $E-K^2/4m$.
The boost results in the multiplicative factor $e^{i\bm{K}\cdot \rvec/2}$ in \eq{cal-nonlocal}
and can be reproduced in the OPE by appropriate operators. The
one-body results in Table \ref{tab1} can be written in the compressed
form
$ \psisdagg \,e^{\rvec \cdot\!\! \arrowpartialvec \!\!/ 2} \,\psis$
to reproduce this factor. In the two-body sector, we can account
for a finite center-of-mass momentum in the OPE  by including the
operator
$\psi_1^\dagger \psi_2^\dagger \psi_2 e^{\rvec \cdot
  \!\!\arrowpartialvec '\!\!/2}\psi_1$ in place of $\cO_{2,4}$, where
$\arrowpartial{i}'=\rarrowpartial{i}+\larrowpartial{i}$.  We can
expand the exponential in small $r$ and define two new
operators up to scaling dimension 6:
\begin{align}
\cO_{2,5}'&=2i\, \psiadagg\psibdagg \psib \arrowpartial{i}' \psia^\text{(ren)}\,,
\nn\\
\cO_{2,6}'
&=-\tfrac{1}{2}\psiadagg\psibdagg \psib \arrowpartial{i}' \arrowpartial{j}' \psia^\text{(ren)}
\,.\nn
\end{align}
Note that the Wilson coefficients are trivial in the sense that, after 
including the appropriate factors from the definitions of $\cO_{2,5}'$
and $\cO_{2,6}'$, we simply use the coefficient $W_{2,4}$.

\subsection{The Momentum Distribution and the Hamiltonian}
Collecting all the one- and two-body matching results and Fourier
transforming to momentum space, we obtain
\begin{widetext}
\begin{align} \label{eq:OPE-k}
\rho_\sigma(\kvec)
&=\int_\Rvec\,\int_\rvec \, e^{i \kvec\cdot \rvec} \langle\psisdagg (\Rvec-\rvec/2)\psis(\Rvec+\rvec/2)\rangle
\nn\\	&= 
\int_\Rvec \Bigg\{
\delta_{\kvec}\, \langle\cO_{1,3}(\Rvec)\rangle
 -i\frac{\nabla_{k_i} \delta_\kvec }{2} \langle \cO_{1,4} (\Rvec)\rangle
- \frac{\nabla_{k_i} \nabla_{k_j} \delta_\kvec}{8}\, \langle \cO_{1,5}^\text{(ren)} (\Rvec)\rangle
 +i \frac{\nabla_{k_i} \nabla_{k_i} \nabla_{k_k} \delta_\kvec }{48}\, \langle\cO_{1,6} (\Rvec)\rangle
\nn \\ &	
\qquad\qquad
+\Bigg(\frac{1}{k^4}+r_s\frac{\delta_\kvec}{8\pi}\Bigg) \, \langle\cO_{C}(\Rvec)\rangle+\frac{1}{k^6}\,\langle \cO_{D}(\Rvec)\rangle
\nn \\ &	\qquad\qquad
+ \Bigg( \frac{\hat{k}_i}{k^5}-r_s\frac{\nabla_{k_i} \delta_\kvec }{32\pi} \Bigg)\langle \cO_{2,5}'(\Rvec) \rangle
+ \Bigg(  \frac{6\hat{k}_i\hat{k}_j-\delta_{ij} }{k^6} +r_s \frac{\nabla_{k_i} \nabla_{k_j} \delta_\kvec}{32\pi}  \Bigg)\langle \cO_{2,6}'(\Rvec) \rangle
\Bigg\}
\,,\end{align}
\end{widetext}
where $\int_\Rvec=\int d^3\Rvec$, $\int_\rvec=\int d^3\rvec$, $\delta_\kvec=(2\pi)^3\delta^{(3)}(\kvec)$, and the unit vector $\hat{k}_i=k_i/k$.
The third line in \eq{OPE-k} corresponds to the contact $C$ and derivative
contact $D$ in \eq{rhoOPE}. On the last line the term with the $1/k^6$ tail  
corresponds to the term $C' \equiv \int_\Rvec\,  \delta_{ij} \langle \cO_{2,6}'(\Rvec) \rangle$ in \eq{rhoOPE}
after an angular average over $\kvec$,
while the term with $1/k^5$ tail vanishes.

For the derivation of the energy relation in 
\eq{Erel} we first write the
Hamiltonian in the absence of an external potential,
using the terms from the Lagrangian in \eqs{L0}{L1} as $H=H_0+H_1$, where
\begin{align}
\label{eq:H01_orig}
H_0 & =\int_\Rvec\Big[\cT(\Rvec) + \frac{\lambda_{0}}{m}\cO_{2,4}(\Rvec)
\Big] \nn\,,\\
H_1 & =\int_\Rvec\Big[\frac{-\rho_{0}}{4}\cO_{2,6}(\Rvec)
+\frac{\delta\lambda_{0}}{m}\cO_{2,4}(\Rvec)\Big]\,.
\end{align}
Then we rewrite the operators in the Hamiltonian 
in terms of the operators that appear in \eq{rhoOPE}:
\begin{align} \label{eq:H0}
 &H_0
= \int_\Rvec\,\bigg[\cT+  \frac{\cO_C }{4\pi m a}-
         \Bigg(\frac{\Lambda}{2\pi^2 m} +\frac{\rho_0 
         \Lambda^3}{3\pi^2 \lambda_0} \Bigg)\,\cO_C \bigg] 
\,,\\
\label{eq:H1}
&H_1
=\int_\Rvec\,\frac{\rho_{0}}{\lambda_{0}^{2}}\cO_{D}
=\int_\Rvec\,\frac{1}{16 \pi m}\Bigg(r_s-\frac{4}{\pi \Lambda} \Bigg)\,\cO_D
\,.\end{align}
The operators $\cT$, $\cO_C$, and $\cO_D$ are
\begin{align} \label{eq:T}
\cT(\Rvec)  &\equiv \frac{1}{2m}\sum_\sigma\,\nabla\psi^\dagger_\sigma \cdot \nabla\psi_\sigma 
\,,\\ \label{eq:OC}
 \cO_C (\Rvec) &\equiv \cO_{2,4}^{(\text{ren})}=\lambda_0^2 \Big(1+\frac{m\rho_0\Lambda^3}{3\pi^2}\Big) \, \cO_{2,4} 
 \,,\\
\label{eq:OD}
\cO_D (\Rvec)&  \equiv   
-\frac{\cO_{2,6}^{(\text{ren})}}{4}=-\frac{\lambda_0^2}{4}
\Big[ \cO_{2,6} -\frac{4\lambda_0 \Lambda^3}{3\pi^2}\cO_{2,4}
\Big]+O(\rho_{0})
\,,\end{align}
where $\lambda_{0}$ and $\rho_0$ are the bare couplings related to the
scattering length $a$ and effective range $r_s$ as shown in
\eqs{lam0}{rho0}.  Note that in \eq{OD} we were able to drop all terms
proportional to $\rho_{0}$ that appear in the definition of 
$\cO_{2,6}^{(\text{ren})}$
because there is already a factor of $\rho_{0}$ multiplying $\cO_{D}$ in 
the finite range part of the Hamiltonian given by \eq{H1},
and terms proportional to $\rho_0^2$
should be dropped since we are
working only up to NLO in the range. We use \eq{rho0} to rewrite $H_1$
into a term proportional to $r_{s}\cO_{D}$ and a term
proportional to $\frac{1}{\Lambda}\cO_{D}$.

The last term in square brackets in \eq{OD} is a counter term which 
subtracts the divergent part of the matrix element
$\langle\mathcal{O}_{D}\rangle$. This performs the same task as the
term proportional to $\delta\lambda_{0}$ in the Lagrangian described
in Sec. \ref{sec:effect-field-theory}.  Furthermore, we see that in
this case $\frac{\rho_{0}}{\lambda_{0}^{2}}\cO_{D}$ and
$\frac{-\rho_{0}}{4\lambda_{0}^{2}}\cO_{2,6}^\text{(ren)}$, using
\eq{ME26-ren}, give the same result in the Hamiltonian. 

The subtracted kinetic term $T^{\text{(sub)}}$ in \eq{Erel} is defined
by absorbing the explicit cutoff dependence from $H_{0}$ and $H_{1}$
into the kinetic term as
\begin{align}
\label{eq:Tsub}
&\langle T^{\text{(sub)}} \rangle
= \int_\Rvec\, \Bigg\langle \cT -\frac{1}{2\pi^2 m}\Bigg[
			\Lambda\Big(1+\tfrac{2m\rho_0\Lambda^2}{3\lambda_0}\Big)\cO_C
			+\frac{\cO_D}{2\Lambda}\Bigg] \Bigg\rangle                        
\,.\end{align}
$T^\text{(sub)}$ contains the terms proportional to 
$\Lambda\cO_{C}\,,\Lambda^3\cO_{C}$ to
subtract the divergent pieces of $\int_\Rvec\,\big\langle\cT\big\rangle$ and 
the term proportional to $\frac{1}{\Lambda}\cO_{D}$, as mentioned above, to remove the remaining $\Lambda$-dependence.

\section{Conclusion}
\label{sec:con}

In this paper we carried out the operator product expansion for the
momentum distribution of a two-component Fermi gas including all terms
that are first order in the effective range. Using the result for the
momentum distribution of a two-particle scattering state, we matched
all operators up to scaling dimension 6 and obtained the corresponding
Wilson coefficients. We used a sharp cutoff in our calculations and
used an EFT framework to include corrections due to a finite effective
range.

The main results of this work are extended universal relations that
contain the previously known contact $C$ 
and the two quantities $C'$ and $D$. Specifically, their sum
 appears as the asymptote of the subleading $(C'+D)/k^6$ tail in the
momentum distribution. 
The derivative contact $D$ alone also appears in universal relations for 
the total energy, its derivative with respect to the effective range
$r_s$, the pressure relation, and the virial theorem as an effective
range correction of the form $r_s \,D$.  Werner and Castin
\cite{2012PhRvA..86a3626W} first found the subleading tail and its
relation to energy $D=16\pi m\, dE/dr_s$, which we reproduced using
the OPE.

While effective range corrections to observables are generally
suppressed by a factor of $r_s/a$, the size of derivative contact
itself is not suppressed in this way in comparison to the size of contact.
For instance, the QMC simulation in Ref. \cite{2012PTEP.2012aA209C}
obtained, in the unitary limit, the density of the derivative contact
$\cD/k_F^6\approx 0.06$, while the contact density is
$\cC/k_F^4\approx 0.11$. In the BEC limit ($a\to 0^+$), the derivative
contact becomes more important because it scales like
$\cD/k_F^6\propto 1/(k_F a)^3$, while the contact scales like
$\cC/k_F^4\propto 1/(k_F a)$.  Our result for the tail of the
momentum distribution, in the absence of a known value for $\cC'$, already 
improves the description of the numerical many-body
calculation for the same quantity for $k>1.5 k_F$ as shown in
\fig{mom}. The subleading $\cD/k^6$ in the tail gives a correction as
large as 20\% near $1.5 k_F$ in unitary limit.

Our results are the first step towards range-corrected universal
relations for other observables such as the single-particle dispersion
relation, structure factors, and RF spectroscopy. In our calculation
we have not considered the 3-body operator that would lead to the
Efimov effect and a $1/k^5$ tail for the large imbalanced mass ratio
$m_1/m_2>13.6$ \cite{Nishida:2010tm} or in a system of three identical
bosons~\cite{Braaten:2011sz}. This would be an interesting extension
of the work presented here.

\section*{Acknowledgements}
We thank J. Carlson and S. Gandolfi for providing their QMC results
and E. Mereghetti for helpful discussions. We thank E. Braaten for
comments and Y. Castin, and F. Werner for their feedback on the manuscript .  
DK would like to thank the University of
Tennessee, Knoxville for hospitality while this work was started. This
work was supported by the U.S. Department of Energy through the Office
of Science, Office of Nuclear Physics under Contract
Nos. DE-AC52-06NA25396, DE-AC05-00OR22725, an Early Career Research
Award, the LANL/LDRD Program, and by the National Science Foundation
under Grant No. PHY-1516077.
\newpage

\begin{appendix}

\section{Vertex Factors and Loop Integrals}
\label{sec:loop_integrals}
In the calculation of the scattering amplitude we encounter the
loop diagrams shown in \fig{Full_Amplitude} which lead to the integrals
$\mathcal{I}_{2n}(E)$:
\begin{align}
\label{eq:I2n}
\mathcal{I}_{2n}(E,\Lambda)&=\int_q 
\frac{i \qvec^{2n}}{q_0-\frac{\qvec^2}{2m}+i\e}\frac{i}{E-q_0-\frac{\qvec^2}{2m}+i\e}
\nn\\&
=-\frac{im\Lambda^{2n+1}}{2(2n+1) \pi^2 }+mE\,\cI_{2n-2}(E,\Lambda)
 \,,\\
\label{eq:I0}
\mathcal{I}_0(E,\Lambda)&
=-\frac{im}{2\pi^{2}}\left(\Lambda+\frac{\sqrt{mE}}{2}\Big[
i\pi +\ln\Big(\tfrac{\Lambda-\sqrt{mE}}{\Lambda+\sqrt{mE} }\Big)
\Big]\right)
\nn\\
&\approx -\frac{im}{2\pi^2}\Big(\Lambda+\frac{i\pi}{2}\sqrt{mE}-\frac{mE}{\Lambda}+\cdots\Big)
\,,\\
\cI_{2}(E,\Lambda)&\approx -\frac{im}{2\pi^2}\Big(\frac{\Lambda^{3}}{3}+mE\Lambda
+\frac{i\pi}{2}(mE)^{3/2}
\nn\\& \hspace{1.7 cm}
-\frac{(mE)^2}{\Lambda}+\cdots\Big)
\,,\end{align}
where the integral symbol is defined as $\int_q=\int\tfrac{d^4q}{(2\pi)^4}$.
We assume $E>0$ because we carry out the matching for the two-atom
scattering state above the threshold.  The result which is valid above
and below the threshold is given in \cite{Braaten:2007nq}. The even
powers of $\qvec$ in the numerator of $\cI_{2n}(E,\Lambda)$, indicated
by the index $2n$, come from the attachment of the part of the off
shell amplitude with explicit momentum dependence to a loop diagram.

Next, we examine the integrals which arise in the one-body operator
loop diagrams. Depending upon the particular operator under consideration,
the vertex factor in the loop integrals changes, and \tab{v1d} below
shows the one-body vertex factors. These factors,
as well as the various $v_{2,\Delta}(q,l)$, can be derived by inserting 
the definition of the nonrelativistic fermion field into the operators 
listed in \eq{operators} and taking the given spatial derivatives.

The loop integral $\IaD{2n}$, for which \tab{v1d} applies, is
from the last Feynman diagram of \fig{One_body} and is given by
\begin{align}\label{eq:I1d}
&\IaD{2n}(E)
=\int_q \frac{i^2\qvec^{2n}\,v_{1,\Delta}(\qvec)   }{[E-q_0-\tfrac{\qvec^2}{2m}+i\e]^2}\frac{i}{q_0-\tfrac{\qvec^2}{2m}+i\e}
\nn\\
&=-i \frac{d}{dE}\Bigg[ \int_q \frac{i\qvec^{2n}\,v_{1,\Delta}(\qvec)   }
					{E-q_0-\tfrac{\qvec^2}{2m}+i\e}\frac{i}{q_0-\tfrac{\qvec^2}{2m}+i\e}\Bigg]
\,.\end{align}

\begin{table}[t] 
\begin{tabular*}{\columnwidth}{@{\extracolsep{\fill}} c c c c c}
\hline
$\Delta$ & 3& 4& 5& 6 \\
\hline
$v_{1,\Delta}(\pvec)$	&	1& $2i p_i$ & $-4 p_i p_j$ & $-8i p_i p_j p_k$\\
\end{tabular*}
\caption{Vertex Factors for 1-body operators of $\Delta=3..6$.}
\label{tab:v1d}
\end{table}
\begin{table}[t] 
\begin{tabular*}{\columnwidth}{@{\extracolsep{\fill}} c c c c}
\hline
$\Delta$ & 4& 5& 6 \\
\hline
$v_{2,\Delta}(p)$	&	1/2& $2i p_i$ & $-4 \delta_{ij} p_i p_j$\\
\end{tabular*}
\caption{Partial vertex factors for two-body operators of dimensions
  $\Delta=4..6$. For the total vertex factor, one must use
  $v_{2,\Delta}(q,l) = v_{2,\Delta}(q)+v_{2,\Delta}(l)$, where $q$ and
  $l$ are the vertex's ingoing and outgoing momentum,
  respectively. The $\delta_{ij}$ comes from the use of $\nabla^2$
  rather than $\partial_i\partial_j$ in $\cO_{2,6}$.}
\label{tab:v2}
\end{table}

The subscript $2n$ in $\cI_{2n}^{(1,\Delta)}(E)$
indicates the number of powers of $\qvec$ in the numerator of the
integral due to the attachment of the off shell amplitude, and the
superscript $(1,\Delta)$ specifies that this integral corresponds to
the insertion of a one-body operator of dimension $\Delta$ into a loop
diagram.  Inserting the factors $v_{1,\Delta}(\qvec)$ given in 
Table \ref{tab:v1d} into \eq{I1d} gives
\begin{align}\label{eq:I13}
\Ia{3}{2n}(E)
&=-i\frac{d \, \cI_{2n} (E)}{dE}
\,,\\ 
\label{eq:I15}
  \Ia{5}{2n} (E)&=i\frac{4\delta_{ij} }{3} \frac{d\, \cI_{2n+2}  (E)}{dE}
\,,\end{align}
where $\IaD{2n}(E)=0$ for even $\Delta=4,6$ because their integrand is
an odd function of $\qvec$.  By using $\cI_{2n}$ given in \eq{I2n}, we
obtain explicit expressions for $\IaD{2n}$.

For two-body operators, the integrals $\IbD{2n}$ represent loop integrals
useful in the diagrammatic calculations of \fig{Two-body}.
\begin{align}\label{eq:I2d}
\IbD{2n}(E)
&= \int_q\,\frac{i\qvec^{2n}\,v_{2,\Delta}(\qvec)   }{q_0-\tfrac{\qvec^2}{2m}+i\e}\frac{i}{E-q_0-\tfrac{\qvec^2}{2m}+i\e}
\,.\end{align}
Inserting the vertex factors $v_{2,\Delta}$ from \tab{v2} into \eq{I2d} gives
\begin{align}\label{eq:I24}
\Ib{4}{2n}(E) &=\frac{\cI_{2n}(E)}{2}
\,,\\ 
\label{eq:26}
\Ib{6}{2n}(E) &=-\frac{4\delta_{ij}}{3} \cI_{2n+2}(E)
\,,\end{align}
while $\Ib{5}{2n}(E)=0$. The even powers of loop momentum in the
integrals again come from the attachment of the momentum-dependent part of
the off shell amplitude to loop diagrams. Additional powers of
momentum may be added to the numerator in \eq{I2d} by the vertex
factors given in \tab{v2}.

The following integrals are for the nonlocal operator diagrams for the LHS of
the momentum distribution.
\begin{align}\label{eq:Irho}
  \cI_{\rho,2n}(E)&=\int_q \frac{i^2 \qvec^{2n} e^{i\qvec\cdot\rvec} }{[q_0-\tfrac{\qvec^2}{2m}+i\e]^2}\frac{i}{E-q_0-\tfrac{\qvec^2}{2m}+i\e}
                    \,,\\\nn
  \cI_{\rho,0}(E)&=-\frac{im^2 }{8\pi \sqrt{mE}}e^{i \sqrt{mE}\,r}+O(1/\Lambda^3)
                   \,,\\\nn
  \cI_{\rho,2}(E)&=-\frac{im^2[ \sqrt{mE}-2i/r] }{8\pi}e^{i \sqrt{mE}\,r}+O(1/\Lambda)
                   \,.
\end{align}
 \\
  \\

\end{appendix}
\newpage
\bibliographystyle{apsrev}

\end{document}